\documentclass[preprint,sort&compress,2013,12pt,5p,times,twocolumn]{elsarticle}
\usepackage[pdftex,bookmarks,colorlinks,breaklinks]{hyperref}  
\usepackage{ascmac,amsmath}
\usepackage{bm}
\usepackage{graphicx,epsfig,epstopdf}
\usepackage{geometry}
\usepackage{lipsum}
\usepackage{lineno}

\makeatletter

\journal{Nuclear Physics A}
\begin{document}
\begin{frontmatter}
\title{Near threshold angular distributions of the $^2$H$(\gamma,\Lambda)$X reaction }

\begin{centering}
\author[focal]{B. Beckford\corref{cor1}}
\cortext[cor1]{Corresponding author, Present address: American Physical Society, One Physics Ellipse, College Park, MD, 20740, USA}
\ead{beckford@aps.org}

\author[rvt3]{P. Byd$\breve{z}$ovsk$\acute{y}$}
\author[focal]{A. Chiba}
\author[focal]{D. Doi}
\author[focal]{T. Fujii}
\author[focal]{Y. Fujii}
\author[focal]{K. Futatsukawa\corref{cor2}}
\cortext[cor2]{Present address: High Energy Accelerator Research Organization, KEK, Tsukuba, Ibaraki 305-0801 Japan}
\author[focal]{T. Gogami}
\author[focal]{O. Hashimoto}
\ead{Deceased 3 February 2012}
\author[rvt2]{Y. C. Han\corref{cor3}}
\author[rvt]{K. Hirose}
\author[focal]{R. Honda}
\author[focal]{K. Hosomi\corref{cor4}}
\cortext[cor4]{Present address: Advanced Science Research Center (ASRC), Japan Atomic Energy Agency, Tokai, Ibaraki, 319-1195, Japan}
\author[rvt]{T. Ishikawa}
\author[focal]{H. Kanda}
\author[focal]{M. Kaneta}
\author[focal]{Y. Kaneko}
\author[focal]{S. Kato}
\author[focal]{D. Kawama}
\author[focal]{C. Kimura}
\author[focal]{S. Kiyokawa}
\author[focal]{T. Koike}
\author[focal]{K. Maeda}
\author[focal]{K. Makabe}
\author[focal]{M. Matsubara}
\author[focal]{K. Miwa}
\author[focal]{S. Nagao}
\author[focal]{S. N. Nakamura}
\author[focal]{A. Okuyama}
\author[focal]{K. Shirotori\corref{cor5}}
\author[focal]{K. Sugihara}
\author[rvt]{K. Suzuki}
\author[focal]{T. Tamae}
\author[focal]{H. Tamura}
\author[focal]{K. Tsukada}
\author[focal]{F. Yamamoto}
\author[focal]{T. O. Yamamoto}
\author[rvt]{Y. Yonemoto}
\author[rvt]{H. Yamazaki}

\end{centering}
 \address[focal]{Department of Physics, Tohoku University, Sendai, 980-8578, Japan}
  \address[rvt]{Research Center for Electron Photon Science, Tohoku University, Sendai, 982-0826, Japan }
 \address[rvt2]{School of Nuclear Science and Technology, Lanzhou University, Lanzhou, 730000, China }
  \address[rvt3]{Nuclear Physics Institute , $\breve{R}$e$\breve{z}$ near Prague, Czech Republic, 25068}
 \date{\today}

\begin{abstract}
A study of the $^2$H$({\gamma},{\Lambda})$X reaction was performed using a tagged photon beam at the Research Center for Electron Photon Science (ELPH), Tohoku University.  The photoproduced $\Lambda$ was measured in the $p{\pi^{-}}$ decay channel by the upgraded Neutral Kaon Spectrometer (NKS2+).  The momentum integrated differential cross section was determined as a function of the scatting angle of ${\Lambda}$ in the laboratory frame for five energy bins.  Our results indicated a peak in the cross section at angles smaller than cos$\theta^{LAB}_{\Lambda}$ = $0.96$.  The experimentally obtained angular distributions were compared to isobar models, Kaon-Maid (KM) and Saclay-Lyon A (SLA), in addition to the composite Regge-plus-resonance (RPR) model. Both  SLA(r$K_{1}K_{\gamma}$ = $-1.4$)  and RPR describe the data quite well in contrast to the KM model, which substantially under predicted the cross section at the most forward angles.  With the anticipated finalized data on ${\Lambda}$ integrated and momentum dependent differential cross sections of $^2$H$({\gamma},{\Lambda})$X~\cite{Kaneta_Beckford}, we present our findings on the angular distributions in this report.
\end{abstract}

\begin{keyword}
strangeness, photoproduction, tagged photon
\end{keyword}
\end{frontmatter}

\section{Introduction}
Experimental and theoretical study of the photo-nuclear production of strangeness has been performed extensively on the proton but most recently the focus has  shifted to the neutron, in order to fully idealize a universal description of the process for all six isospin channels. A lone neutron target is unrealistic, thus the use of a deuteron as a quasi-free neutron target is the most practical option.  At present, there has been  experimental investigation of the elementary kaon photoproduction process  on a proton target via the $p(\gamma, K^+) \Lambda(\Sigma)$ reactions that have measured cross sections and some polarization observables at JLab (CLAS)~\cite{CLAS_data,CLAS_data2,CLAS_data3,CLAS_data4}, Spring-8 (LEPS)~\cite{LEPS_data,LEPS_data2}, ELSA (SAPHIR)~\cite{SAPHIR_data,SAPHIR_data_98}, GRAAL~\cite{GRAAL_data,GRAAL_data2} and MAMI~\cite{Blomqvist}.  
More recently, a few collaborations such as those at ELSA and MAINZ~\cite{bantawa_phd,shende_phd} have used liquid deuterium targets in experiments and have measured neutral kaons in the $p(\gamma, K^0) \Lambda(\Sigma)$ reactions. Prior to the decommissioning in 1999, the SAPHIR group reported on the total and differential cross sections as well as hyperon polarizations in the aforementioned reactions. Their results confirmed that the cross section quickly climbs at 1.1 GeV, plateaus and then declines at photon energies close to 1.45 GeV. The published results in 2004 by SAPHIR~\cite{SAPHIR_data}  for $p({\gamma}, K^{+}) {\Lambda}$, were consistent with the previous SAPHIR data~\cite{SAPHIR_data_98}. They concluded the presence of sizable resonance contributions to the production of $\Lambda$ at threshold energies and also for ($\Lambda ,\Sigma$) at approximately 1.45 GeV. Additionally, they reported that hyperons, ($\Lambda, \Sigma^{0}$), were intensively polarized and produced with opposite signs~\cite{SAPHIR_data}. At the time of publication, the CLAS collaboration contributed the largest data set for these reactions at photon energies of 1.6$-$2.53 GeV and angles of $-0.85$ $\le$ cos${\theta}_{K^{+}}^{CM}$ $\le$ 0.95. Despite the substantial experimental data sets that have been measured for the $p({\gamma}, K^{+}) {\Lambda}$ reaction with CLAS and SAPHIR~\cite{CLAS_data3,CLAS_data,CLAS_data4,SAPHIR_data_98,SAPHIR_data} they are still not adequate to constrain theoretical models and successfully predict the cross section of the unmeasured channels, further motivating this study. 

The aim of this experiment was placed on neutral particle ($K^0$ and $\Lambda$) final states at threshold energies. The production of $\Lambda$ may occur through two channels, $\gamma d\rightarrow K^{0}\Lambda (p)$ and $\gamma d\rightarrow K^{+} \Lambda (n)$. Our experiment was conducted near the production threshold and therefore, it acknowledges that the reaction is significantly less perturbed by higher nucleon resonances, allowing for a simplification in the description of the process and permitting the investigation with less uncertainty. The first preliminary results on the energy dependence of the integrated cross section for forward angles and the momentum dependent differential cross sections was previously reported~\cite{futatsukawa_EPJ}. Our experimental and analysis method was centered on the angular dependence of $\Lambda$ photoproduction at photon beam energies ranging from $0.90 - 1.1$ GeV, providing an original measurement and a means to augment the current understanding of neutral kaon production. Theoretically, descriptions of the reaction have used different approaches that included the use of various form-factors, as well as nucleon and hyperon resonances. The data provided here seeks to be used for constructing a credible model with a reduction in the number of uncertain free parameters, while also being capable of extracting information regarding the excitation spectrum.
\section{The NKS2+ Experiment}
Previous exploration of the $^{12}C({\gamma},K^{+}) {\Lambda}$ and $^2$H$({\gamma},K^{0}) {\Lambda}$p processes measured by the Neutral Kaon Spectrometer (NKS) and NKS2 experimental collaborations yielded encouraging results~\cite{Takahashi,watanabe,Tsukada:2007jy,futatsukawa_EPJ,Beckford_Baryon10}. The first measured the neutral channel and demonstrated the importance of the $n({\gamma},K^0)\Lambda$ reaction~\cite{watanabe}. The second reported the first measurement of neutral kaon photo-production on a neutron~\cite{Tsukada:2007jy}. 
   
  
As an outcome of the NKS experiments, the re-envisioned Neutral Kaon Spectrometer (NKS2) was newly designed and constructed at LNS in 2004, replacing the original version. Its main purpose was to investigate the photoproduction process, particularly the production of neutral strange particles via single $K^{0}$ and $\Lambda$ observation with an acceptance less biased in the forward region compared with the NKS spectrometer. The results influenced an extension of the proposed measurable physical observables, which directly propelled advances in the spectrometer design. The NKS2 has since then undergone an additional upgrade in order to improve the spectrometers acceptance, chiefly in the forward hemisphere. The most recent has been the redesign of the inner detector package, by installing an Inner Hodoscope (IH) and a Vertex Drift Chamber (VDC) and being renamed as the NKS2+.

\subsection{Experimental Setup}
A photon beam was generated from a  carbon wire radiator ($\phi 11\mu$m)  and was guided through a collimator in order to reduce the beam halo, a sweep magnet, and into the NKS2+,  where it was directed to the target located at the center. Moving from the inner most position outwards, was the target, which was surrounded by a Vertex Drift Chamber (VDC), and an Inner Hodoscope, comprised of twenty plastic scintillator segments (IH), which acted as the start signal for time of flight measurements. These detectors were fully enclosed in a Cylindrical Drift Chamber (CDC). All detectors were  located between the poles of a dipole magnet with a 680 mm aperture and a 80 cm diameter. The typical magnetic field was 0.42 Tesla at the center position. 
The target was inserted from top side of the magnet yoke with the cryostat also located on the top of the yoke. The target cell shape was cylindrical (30 mm length and 50 mm $\phi$). The cell window, 40 mm $\phi$ in thickness, was made from a polymide film. 
The target system, adapted for the NKS2+, was able to control the liquefaction of deuterium and sustain it in a liquid phase. During the experiment, the temperature and pressure of liquid deuterium and the residual gas were monitored and recorded to estimate the density.  
The average density of the deuterium target was estimated at 0.172 g/cm$ ^{3}$ with minor ambiguities. This corresponded to an approximated average number of deuteron targets of 0.168 {${\mu}b^{-1}$}.

 \section{Data Analysis}
  \begin{figure}[htb]
	\begin{center}

					\includegraphics*[width=0.90\columnwidth]{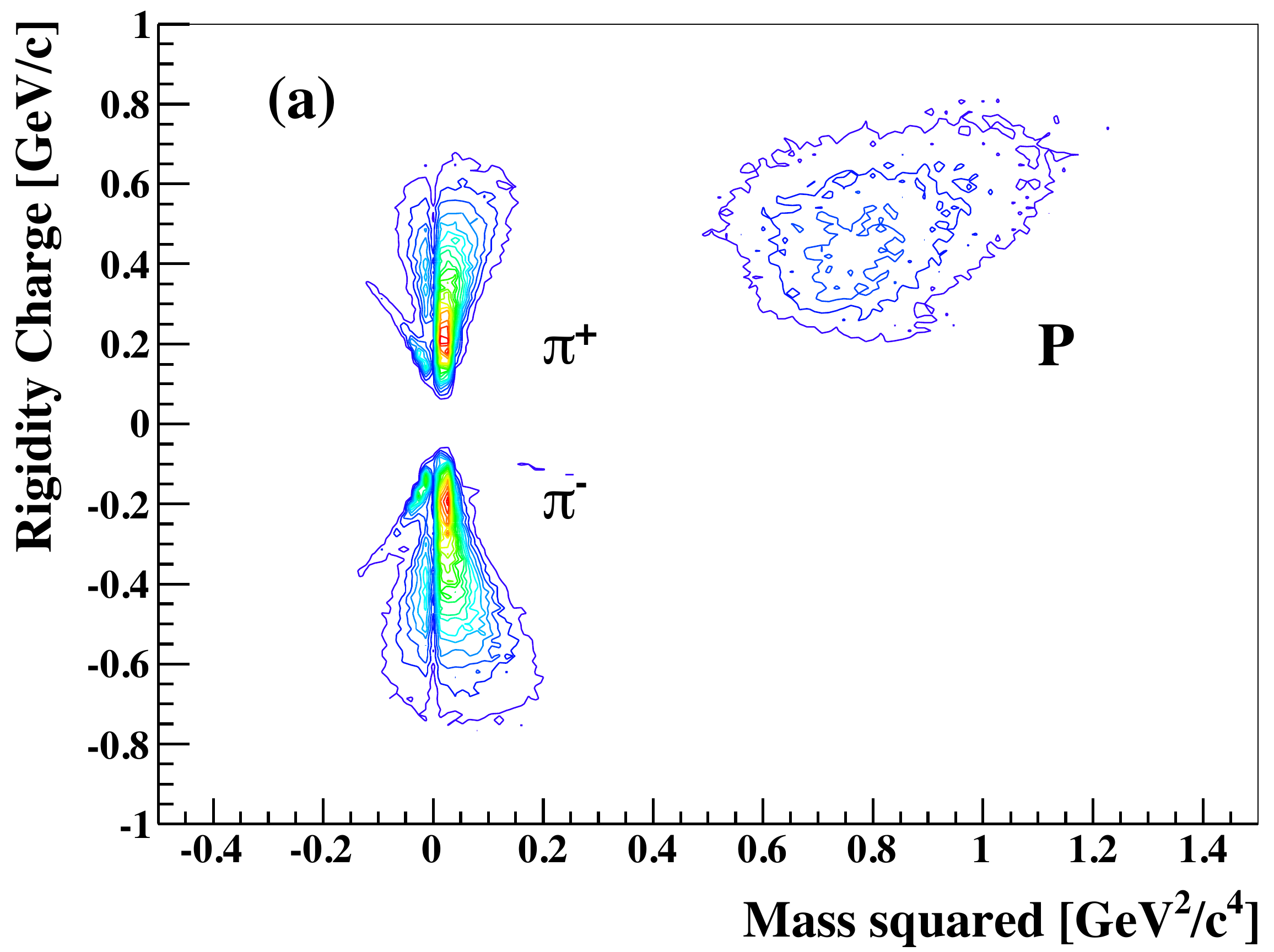}
					\includegraphics*[width=0.90\columnwidth]{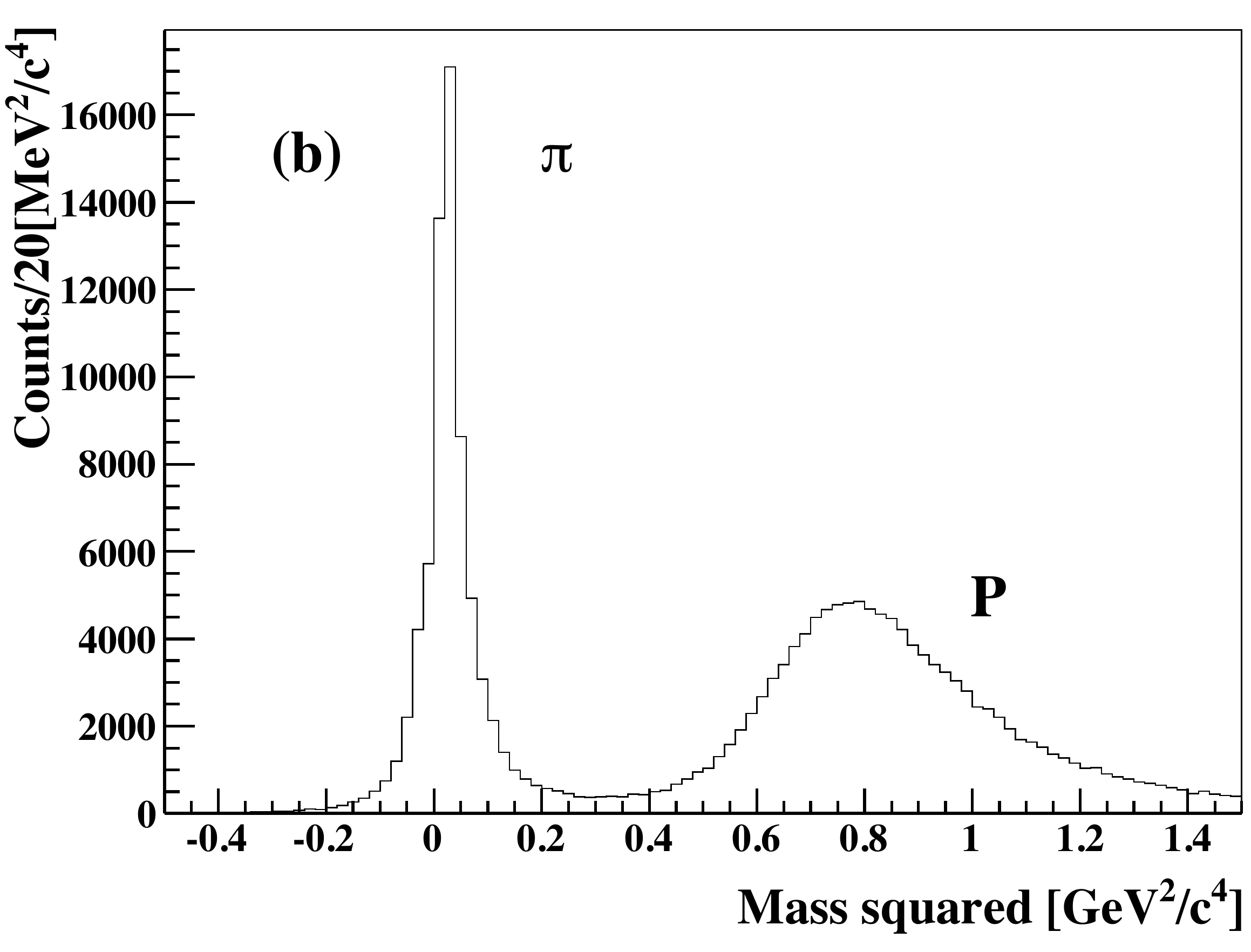}
			\caption{(Color online) Mass squared vs. rigidity ($|\vec{p}|$) multiplied by particle's charge (a), and the projected mass squared distributions (b), are given left to right, where $\pi^{\pm}$ can be identified by the charge.}
  	\vspace{-.2cm}
	\label{fig:mass_squared}
	\end{center}
\end{figure}

The detector commissioning was finished in late 2009 and the experiment discussed here was performed in late 2010. There was additional experimental time scheduled, but that was unfortunately postponed due to  significant damage at ELPH attributed the Tohoku-Pacific Ocean earthquake of 2011/March/11. 
The energy of the photon beam, created via bremsstrahlung, was determined by measuring the scattered electrons from a carbon wire, by the photon tagging system.
The accepted number of events and photons were 0.64 $\times$ 10$^{9} $ and 0.89 $\times$10$^{12}$, respectively. The main trigger a required coincidence signal between the photon tagging counter (Tagger), a two-hit minimum requirement on the IH and OH, along with an absence of a signal from the upstream Electron Veto (EV). A triggered event ensured that the event originated from a real photon and also that the decay particles passed through the length of the IH, drift chambers (VDC $\&$ CDC) and exited the spectrometer via the OH. We employed only the upstream side of EV in the trigger; the downstream side was not included in order to avoid the introduction of a bias in the data set.

\subsection{PID and Event Selection}
The following section details the particle identification and event selection used in the analysis. The normal tagged photon beam rate was about 2 MHz with a trigger rate of roughly 2 kHZ and a 65\% DAQ efficiency. The momenta of the detected particles were determined by the CDC and the species was identified by TOF measurements, with the IH and OH providing the start and stop timing signals respectively. The typical TOF resolution was about 400 ps, which was more than adequate to successfully separate pions and protons at momenta less than 800 MeV/c illustrated in Fig.~\ref{fig:mass_squared}. The direct approach to measuring the photoproduction of strangeness is by reconstructing the invariant mass of the produced particles that contain a hyper charge. A two particle track reconstruction strategy was used in the track analysis of the CDC, which achieved a position resolution between 300 - 400 $\mu$m over its ten layers. A selection requirement was placed on the opening angle between reconstructed particle tracks equivalent to $-0.9$ $\le$ cos${\theta}_{oa}$ $\le$ 0.9 in order to reduce the $e^{+}e^{-}$ background in the data set. The produced $\Lambda$ was detected in the $p{\pi^{-}}$ decay channel. The measured momentum multiplied by the respective particle's charge as a function of squared mass and the projected squared mass distributions are given in Fig.~\ref{fig:mass_squared}. The ${\pi}$ species was identified by its charge and a clear separation between protons and pions is seen, demonstrating PID capability of the NKS2+. The lifetime of $\Lambda (c\tau=$7.89 cm) suggests that the decay take places away from the primary interaction vertex inside the target. Consequently, a selection outside the target was specified to reject particles originating from the principal reaction vertex. 
\begin{figure*}[htb]
	\begin{center}
									\includegraphics*[width=0.90\columnwidth]{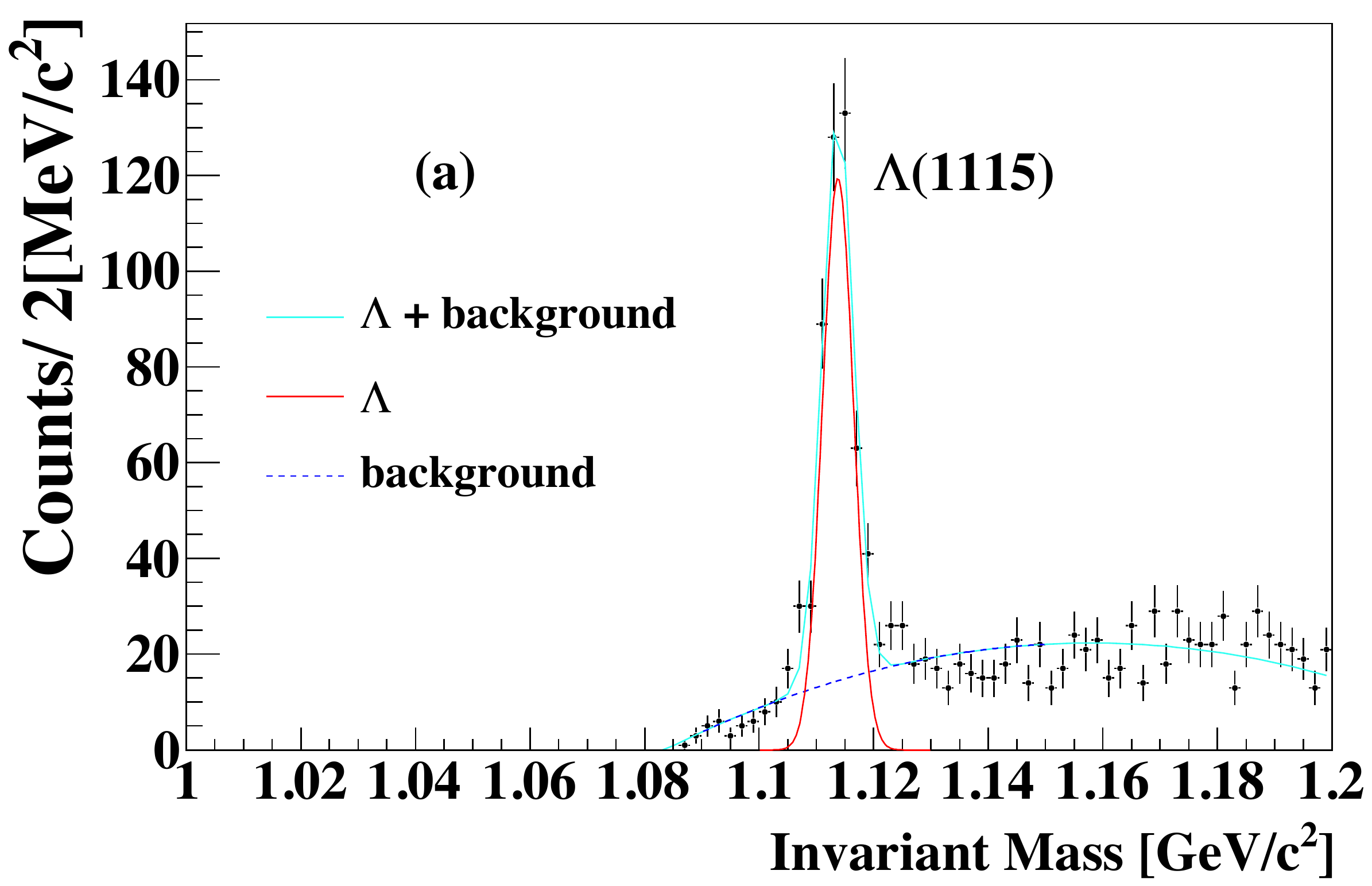}
									\includegraphics*[width=0.95\columnwidth]{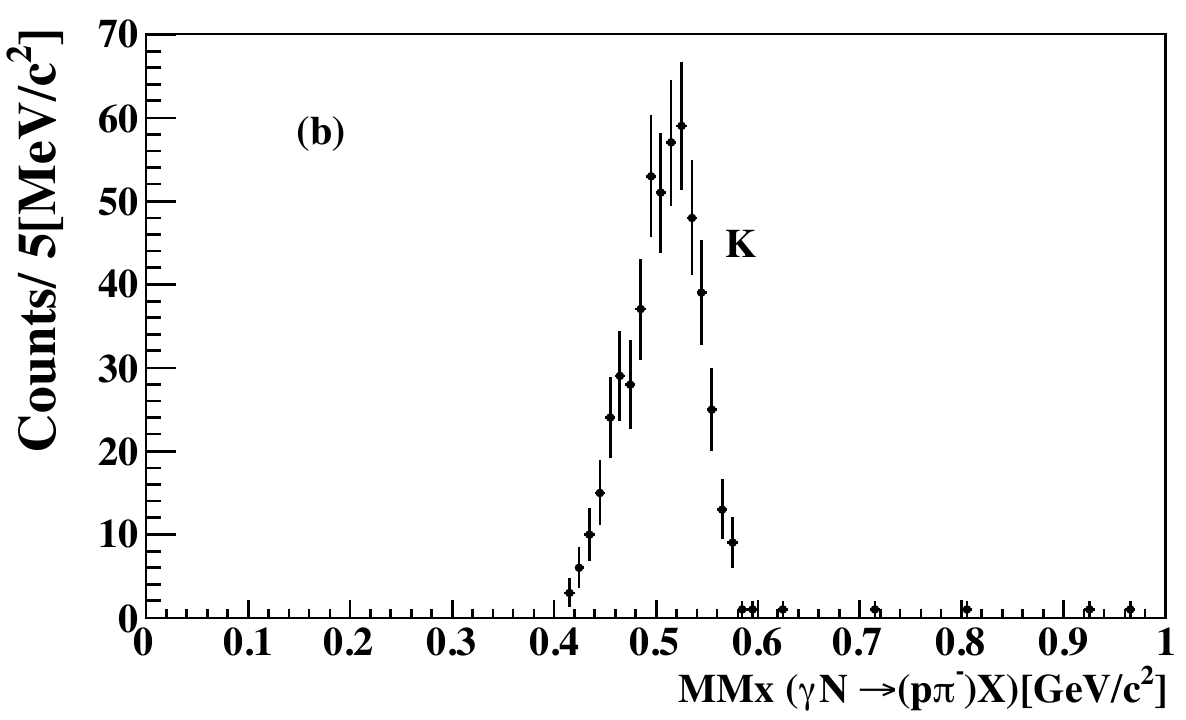}
			\caption{(Color online) Reconstructed invariant mass, p$ {\pi}^{-}$(a) and missing mass, ${\gamma}$N $\rightarrow$p$ {\pi}^{-}$X (b), distributions  are given in the figure left to right respectively. The missing mass distribution from a neutron, assumed at rest within the deuteron nucleus was obtained for events in which a selection was placed on the invariant mass distribution between $1.105 - 1.125$ GeV/$c^2$.}
	\label{fig:im_fitted_E_all_E1_E2}
	\end{center}
\end{figure*} 

\subsection{Invariant Mass Spectra}
Various cuts, but not limited to, timing, $\chi^{2}$ and kinematical requirements were applied to the raw invariant mass to extract a clear $\Lambda$ peak from the continuum, shown in Fig.~\ref{fig:im_fitted_E_all_E1_E2}. The observed $\Lambda$ (roughly 400 events) was detected in the $p{\pi^{-}}$ decay channel with a corresponding (Gaussian) width of 2.87 $\pm$ 0.19 MeV/c (${\sigma}$) and a signal to background (S/B) of 2:1. By placing a tight event selection requirement on the $p{\pi^{-}}$(${\Lambda}$) invariant mass,  between $1.105 - 1.125$ GeV/$c^2$, the missing kaon mass distribution, gated for a neutron at rest within the deuteron nucleus was obtained. The missing mass (Gaussian) width was 31 $\pm$ 2.0 MeV/c$^2$ ${\sigma}$) with a mean value of 504 $\pm$ 2.0 MeV/c$^2$ resulting from detector resolutions. The spread in the missing mass spectrum distribution is associated with the Fermi motion within the deuteron. The generated missing mass was derived from an inclusive measurement of p${\pi^-}$, therefore, the distribution has contributions from the $K^{+} \Lambda$ and $K^{0}\Lambda $ processes. The missing mass technique confirms that the event of interest, $^2$H$({\gamma},{\Lambda}$)KN, was measured. A small contribution to the yielded $\Lambda$ events originates from the $\Sigma^{0}\rightarrow \Lambda \gamma$ channel, but was considered to be minor and could not be separated as the NKS+ lacked a photon detector. 
				

\begin{figure*}[ht]
	\begin{center}

						\includegraphics*[width=0.95\columnwidth]{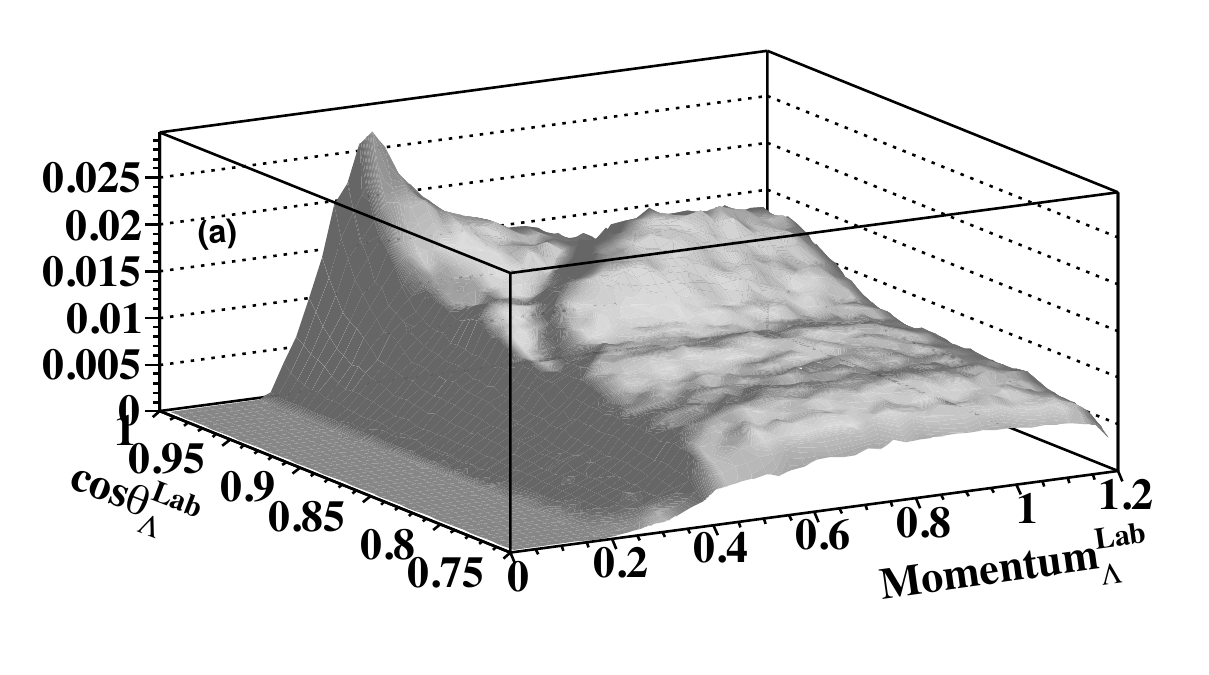}
						\includegraphics*[width=0.95\columnwidth]{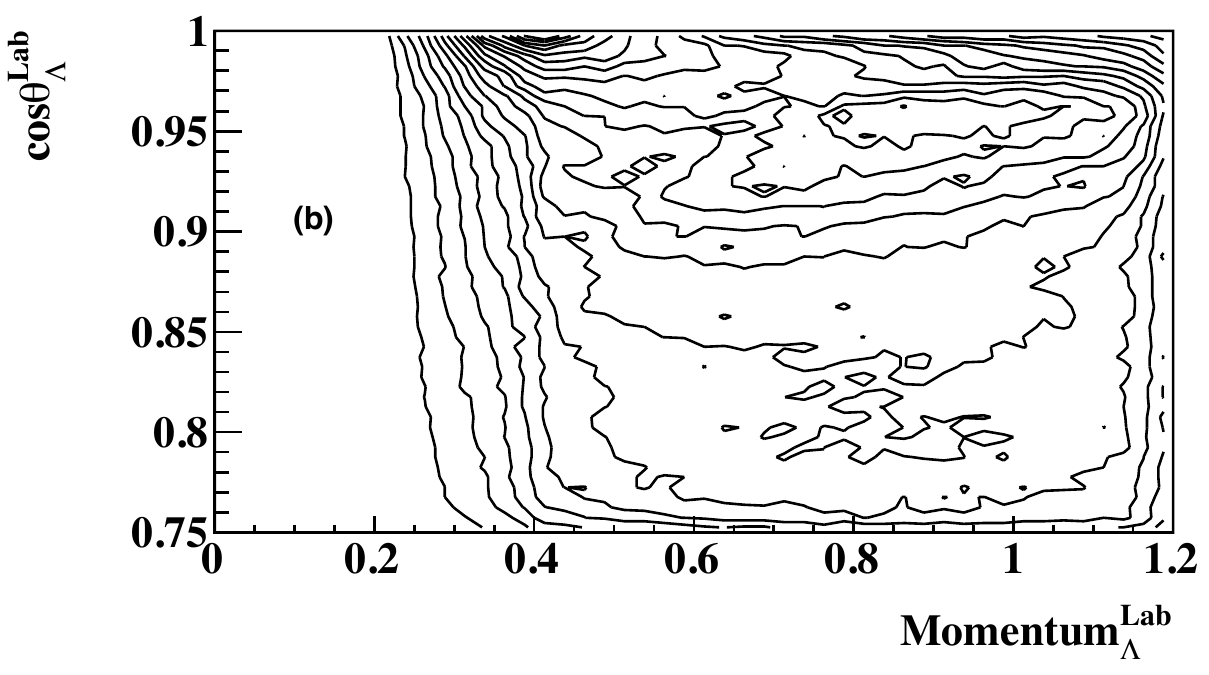}
			\caption{NKS2+ acceptance for $\Lambda \rightarrow p\pi^{-}$ decay. The right figure represents the contour
plot by 0.5\%}
	 \label{fig:lambda_acceptance_experimental}
	\end{center}
\end{figure*} 


The acceptance was estimated by monte-carlo as the ratio of the triggered or reconstructed ${\Lambda}$ events, including the efficiencies of various detector components, to the simulated generated events in the same kinematical bins. It was found by the following: $ {\varepsilon}^{\Lambda}_{\mathrm{accept}}(p,cos{\theta})  = \frac{N_{\mathrm{trig}}} {N_{\mathrm{gen}}}$, where $N_\mathrm{trig}$ was the sum of the ${\Lambda}$ histogram events within the appropriate kinematical range and $N_\mathrm{gen}$ was the number of ${\Lambda}$ histogram events generated isotropically in the lab frame.  In Fig.~\ref{fig:lambda_acceptance_experimental}, the acceptance map for $\Lambda \rightarrow p\pi^{-}$ decay is given; the acceptance is estimated to be at most 0.025.

\section{Results and Discussion}
In this section, the differential cross sections of ${\Lambda}$ as a function of scattering angle after background subtraction are presented for five incident photon energy bins and were compared to two isobar models and the Regge-plus-resonace model calculations.
\subsection{Results}
The differential cross section for the inclusive measurement of  ${\Lambda}$ can be calculated from experimentally measured variables as:

  \begin{equation}
  		\begin{split}
  		  \frac{d{\sigma}}{dpd{\Omega}}= \\ 
		  \frac{N^{\Lambda}_{\mathrm{yield}}(p,cos{\theta})}  {N_{{\gamma}} N_{\mathrm{target}} {\varepsilon}^{\Lambda}_{\mathrm{accept}}(p,cos{\theta}) {\zeta}^{\Lambda} {\varepsilon}^{\Lambda}_{\mathrm{GE}} {\varepsilon}^{\Lambda}_{\mathrm{specific}} 2{\pi}d(cos{\theta})}\\
   			 \label{eq:lambda_diff_cross_section}
   		 \end{split}
  \end{equation}

where $N^{\Lambda}_{\mathrm{yield}}$ was the yield of ${\Lambda}$ events, $N_{\gamma}$ was the number of incident photons on the target, and $N_{\mathrm{target}}$ was the number of target neutrons. The branching ratio for the decay mode, ${\Lambda}$ $\rightarrow p\pi^{-}$ (63.9\%), was represented by ${\zeta}^{\Lambda}$ and the acceptance was ${\varepsilon}^{\Lambda}_{\mathrm{accept}}(p,cos{\theta})$. The term ${\varepsilon}^{\Lambda}_{\mathrm{GE}}$ referred to the efficiencies that are universal to the spectrometer, and ${\varepsilon}^{\Lambda}_{\mathrm{specific}}$ denoted the efficiencies that were intrinsic to the inclusive ${\Lambda}$ measurement.  The efficiency of the accepted p$\pi^-$ invariant mass(${\varepsilon}_{\mathrm{IM}}$) was estimated to be 98.9\% $\pm$ 2\%. The efficiency of the number of hits required to reconstruct a track(${\varepsilon}_{\mathrm{hits}}$), was calculated as 98.3\% $\pm$ 2\%. The uncertainty of the number of photons irradiated on the target which was proportional to the number of recorded scalar hits in TagF (N$_{\gamma}$) was straightforwardly found as the square root of the counts $\sqrt{ N_{\gamma}}$. 

The dominance of the $^2$H$({\gamma},{\Lambda})K^{0}$ reaction, for the SLA calculation increases to more than twice that of $^2$H$({\gamma},{\Lambda})K^{+}$ with extending photon beam energies at angles of cos$\theta^{LAB}_{\Lambda} \ge$0.95 in comparison to KM. The disparity in the contribution of each process arises from the resonance content of each model's approach to Born term suppression. SLA incorporates hyperon resonances to assist in curbing of large non-physical Born term contributions, while the KM model only relies on nucleon resonances. In the $K^{0}\Lambda$ channel, the importance of the $r{K_1K_\gamma}$ parameter is remarkably dissimilar for both models. Previous calculations of the momentum integrated angular cross sections for the SLA(r$K_{1}K_{\gamma}$ = $-2.0$) were reported but were based on fits to published NKS2 data that was later amended~\cite{Bydzovsky_2009,Tsukada:2007jy}, resulting in a newly acquired parameter of SLA(r$K_1K_{\gamma}$ = $-1.4$). The ratio of the photo coupling constants in the ${t}$-channel for the neutral and charged channels are related to the helicity amplitudes, such that
 \begin{equation}
 	\begin{split}
 rK_1K_{\gamma} = \frac{g{K_1^{0}}K_{\gamma}^{0}} {g{K_1^{+}K_{\gamma}^{+}}}
   = -\sqrt{\frac{(\Gamma_{K_{1}^{0}\rightarrow K_{\gamma}^{0}})}{\Gamma_{K_{1}^{+}\rightarrow K_{\gamma}^{+}}}} 
  	\end{split}
  \end{equation}
 Transition moments correspond to decay widths of the meson, where quark model calculations assist in restricting the values of the K$^{*}$(829) meson. The decay widths for the K$_{1}$ meson are undetermined and are fixed from global fits to the data for KM, but left as a free fitting parameter in SLA. More recent SLA calculations with the parameter (r$K_1K_{\gamma}$ = $-1.4$) were used in our comparisons~\cite{ Bydzovsky_private}. The $r{K_1K_\gamma}$ parameter in the KM framework was determined by fitting $K^{0}\Sigma$ data. Therefore, the conclusion can be drawn that the higher sensitivity of the K$1$ meson was afforded to SLA but seems to modestly affect the KM predictions~\cite{Bydzovsky_private}

The background subtracted differential cross sections after acceptance corrections are given in Fig.~\ref{fig:angular_distributions_KM_SLA_RPR_comparison} for the $^2$H$({\gamma},{\Lambda})$X reaction in the energy regions of $0.95-1.00, 1.02-1.04, 1.04-1.06, 1.06-1.08$ GeV, presented in figures (a)-(e) respectively. The horizontal and vertical errors bars correspond to the bin width and statistical errors. The data revealed that the cross sections were concentrated at laboratory angles of cos${\theta}_{\Lambda}^{LAB}$$\ge$ 0.85, therefore, the ability to measure the total cross of $^2$H$({\gamma},{\Lambda})$X is feasible with a spectrometer capable of measuring these small angles as predicted~\cite{Bydzovsky_2009}. 

\subsection{Errors}
One of the main contributions to the systematic errors was the estimated background under the reconstructed invariant mass given in Fig.~\ref{fig:im_fitted_E_all_E1_E2}. We approximated the background by a variety of methods including fitting the distribution with combinations of gaussian and polynomial functions of increasing degrees. The side-band method was also evaluated as a possible best description of the background. The spread in the obtained background values for the all approximated approaches was used a good measure of the uncertainty and was considered to contribute to the lion's share of systematic error in the calculated cross section.  Additional systematic uncertainties were introduced by the  kinematical cuts, in the determination of the number of target deuterons, and incident photons. We estimated a 3\% systematic error from the opening angle cut and less than 2\% for all the additional stipulations. by monte-carlo. The total systematic error was estimated to be roughly 15\%. The systematic error is shown in Fig.~\ref{fig:angular_distributions_KM_SLA_RPR_comparison} as the shaded regions on the data points.
  
\subsection{Discussion}
 \begin{figure*}[ht]
 \begin{center}

                          \includegraphics[width=1.99\columnwidth]{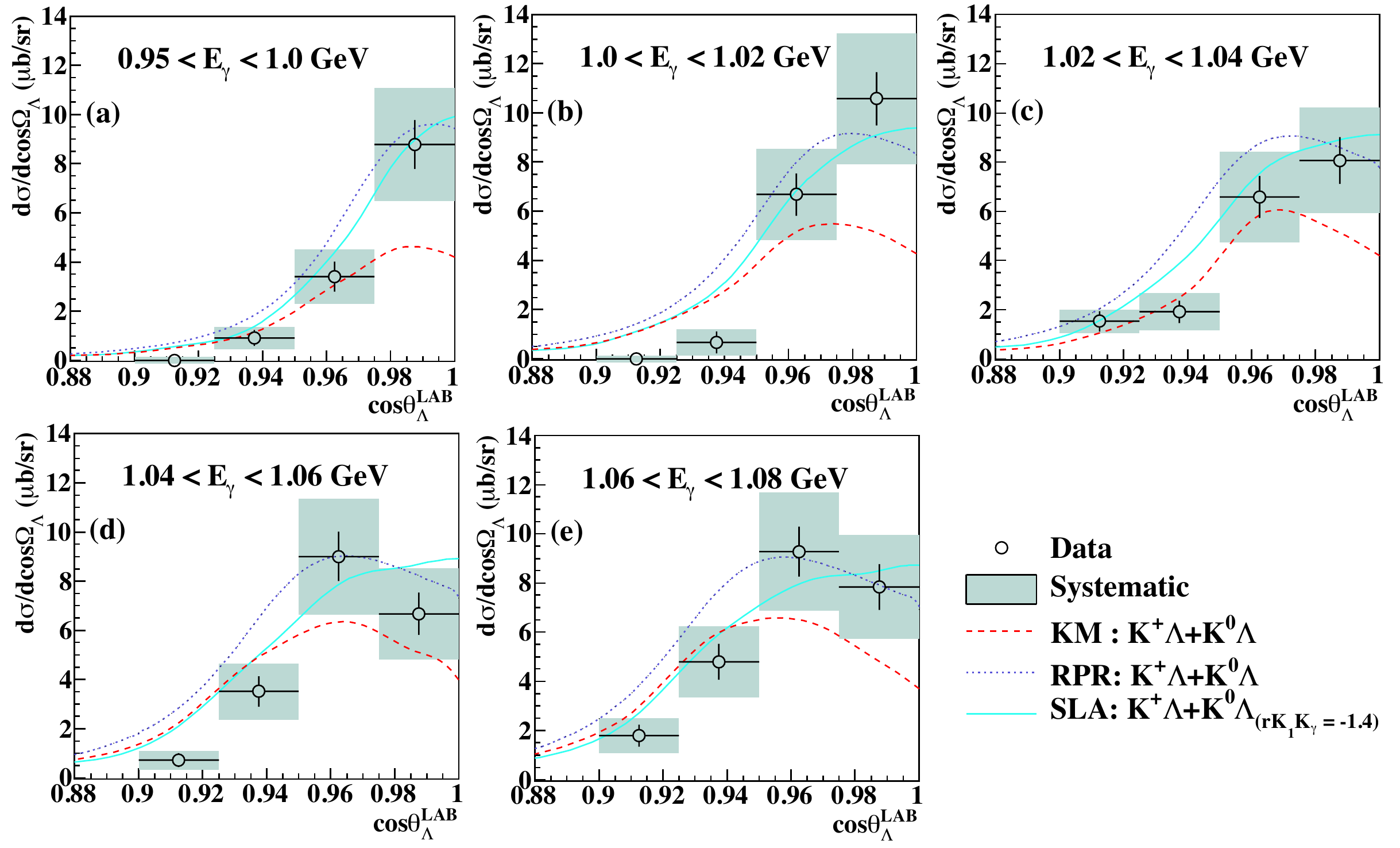}
                                                        
\caption{(Color online) Background subtracted differential cross section as a function of the scattering angle (cos${\theta}_{\Lambda}^{LAB}$), compared with theoretical predictions of the $^{2}$H$({\gamma},{\Lambda})KN$ reaction, calculated by the KM, SLA, and RPR models. Photon energy bins are indicated in the figures where the bin width and statistical errors are given as the horizontal and vertical errors respectively. The systematic error was estimated to be roughly 15\% given as the shaded regions on the data points.}
       \label{fig:angular_distributions_KM_SLA_RPR_comparison}
       \end{center}
  \end{figure*}
  
 The angle dependent distributions were compared to the calculated predictions of effective Lagrangian (isobar) models, Kaon-Maid (KM)~\cite{Benn_Mart} and Saclay-Lyon A (SLA)~\cite{SLA,SLA_MIZU}, as well as the hybrid Regge-plus-resonance (RPR) model~\cite{RPR_kaon_deuteron_AIP,RPR-Model, RPR_jul_2012}. The measured cross sections are plotted alongside the theoretical predictions of the summation of the contributing cross sections of the $^2$H$({\gamma},K^{+}) {\Lambda}$n and $^2$H$({\gamma},K^{0}) {\Lambda}$p reactions in the laboratory frame as a function of cos$\theta^{LAB}_{\Lambda}$ for KM, SLA, and RPR in Fig.~\ref{fig:angular_distributions_KM_SLA_RPR_comparison}, where the curves for KM, SLA, and RPR are drawn as the dotted, dashed, and solid curves respectively. 
In the reported integrated energy bins, RPR accomplished a good description of the experimental results with agreements within the statistical error.  In the lower energy bins, $0.95 - 1.00$ GeV (a), $1.00 - 1.02$ GeV (b), the model over predicted the angular cross section by approximately $20 - 30$\%.  In the higher energy bins, it is evident that the model over estimated the cross sections by around $10 - 15$\% for cos$\theta^{LAB}_{\Lambda} \ge$ 0.95.  The strongest accord between the data and the RPR calculations existed in the $1.02 - 1.04$ GeV (c) energy bin. The RPR and SLA(r$K_1K_{\gamma}$ = $-1.4$) models did a reasonable job of describing the general trend of the data. SLA  however to described the apparent peak at cos$\theta^{LAB}_{\Lambda}$ = 0.96, and demonstrated better reproduction of the data than RPR at higher energies and smaller angles. KM did not describe the data well, most significantly in the extreme forward region at angles smaller than cos${\theta}_{\Lambda}^{LAB}\ge$ 0.96. The amplitudes for the SLA predictions were significantly larger than those of KM for laboratory angles of cos$\theta^{LAB}_{\Lambda}\ge$0.95. Unlike the results of the KM and RPR, the $^2$H$({\gamma},{\Lambda}) K^{0}p$ process is the primary contributor in the SLA framework. 

\subsection{Conclusion}
We have measured the angular dependence of $\Lambda$ photoproduction in the threshold region with the upgraded Neutral Kaon Spectrometer 2 (NKS2+) utilizing tagged photons on a liquid deuterium target. Cross sections were dominant at small scattering angles roughly equal to cos$\theta^{LAB}_{\Lambda}$ = $0.96$, and were reduced at most forward angles at higher energies. Our findings were supported by calculations of SLA with r$K_1K_{\gamma}$=$-1.4$ providing a decisive means to further contain model parameters. RPR predictions were within the given statistical errors, however their calculations exhibited an over estimation of the $K^{+}\Lambda$ contribution to the cross sections. The most significant conclusion was the demonstrated ability of the NKS2+ to roughly measure the total cross section of $\Lambda$ in the reaction.\\

\section{Acknowledgements} 
  \vspace{-.2cm}
This worknwas partly supported by the JSPS Grant-in-Aid for Creative research program (16GS0201), the JSPS Core-to-Core program (21002), and
the JSPS Grant-in-Aid for Scientific Research (23540334). The NKS2 collaboration expresses its thanks to the accelerator group of ELPH, Tohoku University. The theoretical calculations for the RPR-2007 model were performed by P. Vancraeyvald~\cite{RPR_private}.

\bibliographystyle{unsrt}

\end{document}